\documentclass[12pt]{article}
\usepackage{epsfig}
\usepackage{graphicx}
\usepackage{amsmath}
\usepackage{amssymb}
\usepackage{slashed}
\newcommand{\beq}{\begin{equation}}
\newcommand{\eeq}{\end{equation}}
\newcommand{\bea}{\begin{eqnarray}}
\newcommand{\eea}{\end{eqnarray}}
\newcommand{\bem}{\begin{multline}}
\newcommand{\eem}{\end{multline}}
\newcommand{\beg}{\begin{gather}}
\newcommand{\eeg}{\end{gather}}

\def\eq#1{{Eq.~(\ref{#1})}}
\def\fig#1{{Fig.~\ref{#1}}}
\newcommand{\ben}{\begin{eqnarray*}}
\newcommand{\een}{\end{eqnarray*}}

\begin{document}
\title{{\bf Single Inclusive Hadron Production at RHIC and the LHC from the Color Glass Condensate \\[1.5cm] }}
\author{
{\bf Javier L. Albacete\thanks{e-mail: javier.lopez-albacete@cea.fr}\hspace{0.2cm} and 
 Cyrille Marquet\thanks{e-mail: cyrille.marquet@cea.fr}}
\\[0.5cm] {\it Institut de Physique Th\'eorique - CEA/Saclay}, \\
{\it 91191 Gif-sur-Yvette cedex, France.}\\}
\maketitle

\begin{abstract}

Using the unintegrated gluon distribution obtained from numerical simulations of the Balitsky-Kovchegov equation with running coupling, we obtain a very good description of RHIC data on single inclusive hadron production at forward rapidities in both p+p and d+Au collisions. No K-factors are needed for charged hadrons, whereas for pion production a rapidity independent K-factor of order 1/3 is needed. Extrapolating to LHC energies, we calculate nuclear modification factors for light hadrons in p+Pb collision, as well as the contribution of initial state effects to the suppression of the nuclear modification factor in Pb+Pb collisions.

\end{abstract}




\section{Introduction}

The suppression of particle production at forward rapidities in d+Au collisions compared to p+p collisions, experimentally observed at RHIC \cite{Arsene:2004ux,Adams:2006uz}, constitutes one of the most compelling indications for the presence of non-linear QCD evolution effects in presently available data. The appropriate framework to study the nuclear wave function in this non-linear QCD saturation regime is the Color Glass Condensate (CGC), see e.g. the reviews
\cite{Iancu:2003xm,Weigert:2005us} and references therein. The CGC is endowed with a set of
non-linear pQCD evolution equations, the JIMWLK equations, which  in the large-$N_c$ limit reduce to the Balitsky-Kovchegov (BK) equation \cite{Balitsky:1996ub,Kovchegov:1999yj}. The BK-JIMWLK equations can be interpreted as a renormalization group equation for the Bjorken-$x$ evolution of the unintegrated gluon distribution, and more generally of $n$-point correlators averaged over the nuclear wave function, in which both linear radiative processes and non-linear {\it recombination} effects are included. 

Indeed, the observed reduction of the forward hadron yield in d+Au collisions was predicted, 
albeit at a qualitative level, in \cite{Kharzeev:2003wz,Albacete:2003iq}, where it was directly 
related to the {\it shadowing} built in the wave function of the gold nucleus due to the enhanced 
role of non-linear effects in its evolution towards larger rapidities (smaller $x$). 
Later on, a better quantitative description of the d+Au forward hadron yields was achieved in the 
CGC calculations of \cite{Kharzeev:2004yx,Dumitru:2005kb,Goncalves:2006yt,Boer:2007ug}. 
These works relied on the use of models for the unintegrated gluon distribution of the gold 
nucleus inspired by approximate solutions of the BK equation. Relevant dynamical features in 
these models where either taken from analyses of lepton-proton scattering data or directly fitted to data, such as the {\it anomalous dimension} or the rapidity dependence of the saturation momentum $Q_s,$ the scale below which non-linear effects become important.
More detailed analytical and phenomenological analyses of the corresponding nuclear modification factors were carried out in \cite{Baier:2005dz} and \cite{Armesto:2004ud,Tuchin:2007pf} respectively.

The reason why the BK-JIMWLK equations, the most robust theoretical tool available to describe the small-$x$ dependence of the nuclear wave function, have not been directly used in phenomenological studies, is that they were originally derived at leading-logarithmic accuracy only. It was quickly understood that this was not good enough: in analytical \cite{Iancu:2002tr,Mueller:2002zm,Munier:2003vc,Munier:2003sj} and numerical \cite{Armesto:2001fa,Golec-Biernat:2001if,Albacete:2004gw,Marquet:2005zf} studies of the original leading-order (LO) BK equation, the growth of the saturation scale was determined to be $Q_s^2\sim x^{-\lambda_{LO}}$, with $\lambda_{LO}\approx
4.8\,\frac{N_c}{\pi}\,\alpha_s,$ incompatible with the phenomenology of lepton-hadron scattering which demands $Q_s^2\sim x^{-0.2\div 0.3}$. Moreover there were hints that higher-order corrections would restore the compatibility of these values \cite{Triantafyllopoulos:2002nz}.

However, such insufficiency of the theory has been (at least partially) fixed through the recent 
calculation of the next-to-leading order (NLO) evolution equation. First, running-coupling corrections to the LO BK-JIMWLK kernel were derived in \cite{Gardi:2006rp,Kovchegov:2006vj,Balitsky:2006wa}. Then, the full NLO BK equation was obtained \cite{Balitsky:2008zz}. As demonstrated in \cite{Albacete:2007yr}, one can account for most of the higher-order contribution with running-coupling corrections only. Importantly, higher order corrections bring the evolution speed, $\lambda$, down to values compatible with experimental data, among other interesting dynamical effects, thus narrowing the gap between theory and data.

Indeed, first steps in promoting the BK equation with running-coupling corrections (referred to as rcBK henceforth) to an operational phenomenological tool have been taken in Refs.~\cite{Albacete:2007sm,Albacete:2009fh,Betemps:2009ie,Dusling:2009ni}. In \cite{Albacete:2007sm}, a good description of the rapidity and collision energy dependence of the hadron multiplicities in Au+Au collisions measured at RHIC was achieved. Ref.~\cite{Albacete:2009fh} demonstrated the ability of the rcBK equation to account for the small-$x$ behavior of the total ($F_2$) and longitudinal ($F_L$) structure functions measured in
e+p scattering experiments. Then it was shown in \cite{Betemps:2009ie} that the proton scattering amplitude fitted to data in \cite{Albacete:2009fh} allows a good simultaneous description of both the proton diffractive structure function measured at HERA and of the forward hadron yields measured in p+p collisions at RHIC. Finally, in \cite{Dusling:2009ni} a good description of the few nuclear structure functions known at small $x$ from e+A experiments was obtained.

Together, these works yield a consistent picture that present experiments can probe the non-linear part of the hadronic and nuclear wave functions at small $x$, and that they can be successfully described by the CGC effective theory of QCD at high energies.
In this work we provide a good description of the single inclusive hadron (charged hadron and neutral pions) yields measured in p+p and d+Au collisions at RHIC at forward rapidities ($y_h>2$), with unintegrated gluon distributions obtained from the rcBK equation. We also extrapolate our results to LHC energies and predict forward particle production in p+p, p+Pb collisions, that we present through nuclear modification factors for light hadrons. In the case of Pb+Pb collisions, we are able to give the contribution of initial state effects to the suppression of the nuclear modification factor.

\section{Inclusive hadron spectra in d+Au collisions at RHIC}

According to Ref.~\cite{Dumitru:2005gt}, the differential cross section for forward hadron production in proton-nucleus collisions is given by 
\begin{eqnarray}
\frac{dN_h}{dy_h\,d^2p_t}=\frac{K}{(2\pi)^2}\sum_{q}\int_{x_F}^1\,\frac{dz}{z^2}\, \left[x_1f_{q\,/\,p}
(x_1,p_t^2)\,\tilde{N}_F\left(x_2,\frac{p_t}{z}\right)\,D_{h\,/\,q}(z,p_t^2)\right.\nonumber\\ 
+\left. x_1f_{g\,/\,p}(x_1,p_t^2)\,\tilde{N}_A\left(x_2,\frac{p_t}{z}\right)\,D_{h\,/\,g}(z,p_t^2)\right]
\label{hyb}\,,
\end{eqnarray}
where $p_t$ and $y_h$ are the transverse momentum and rapidity of the produced hadron, and
$f_{i/p}$ and $D_{h/i}$ refer to the parton distribution function of the incoming proton and to the final-state hadron fragmentation function respectively. Here we will use the CTEQ6 NLO
p.d.f's \cite{Pumplin:2002vw} and the DSS NLO fragmentation functions
\cite{deFlorian:2007aj,deFlorian:2007hc}. In writing \eq{hyb} we have assumed that the factorization and fragmentation scales are both equal to the transverse momentum of the produced hadron. For light hadron production discussed here, the difference between the rapidity and pseudo-rapidity, $\eta_{h}$, of the produced hadron can be neglected, yielding the following kinematics:
$x_F=\sqrt{m_h^2+p_t^2}/\sqrt{s_{NN}}\,\exp{(\eta_h)}\approx p_t/\sqrt{s_{NN}}\,\exp{(y_h)}$, $x_1=x_F/z$ and $x_2=x_1\exp{(-2y_h)},$ with $\sqrt{s_{NN}}$ the collision energy per nucleon.
Finally, the unintegrated gluon distributions (udg's) $\tilde{N}_{F(A)}$ describe the scattering of a {\it hard} valence quark (gluon) from the projectile on the {\it saturated} small-$x$ glue of the target, either a nucleus or a proton. In order to avoid contamination from
large(small)-$x$ effects in the target (projectile), we will restrict ourselves to the study of the forward region $y_h\gtrapprox2$ both at RHIC and LHC energies, such that $x_1\gg x_0$ and $x_2\ll x_0$, where $x_0$ is the $x$-value where the small-$x$ evolution starts (see below). 
Similar to previous approaches, we allow the possibility of a K-factor to absorb the effect of higher order corrections. For instance there is no $\alpha_s$-order term in \eq{hyb}, we shall only implement running-coupling corrections in the $x_2$ evolution of $\tilde{N}_{F(A)}$, but in principle they also affect the cross section \cite{Kovchegov:2007vf}. 

The udg's $\tilde{N}_{F(A)}$ are given by the two-dimensional Fourier transform of the imaginary part of the forward dipole-target scattering amplitude in the fundamental (F) or adjoint (A) representation, $\mathcal{N}_{F(A)}$, respectively:
\begin{equation}
\tilde{N}_{F(A)}(x,k)=\int d^2{\bf r}\,e^{-i{\bf k}\cdot{\bf r}}\left[1-\mathcal{N}_{F(A)}(r,Y\!=\!\ln(x_0/x))\right],
\end{equation}
where $r$ is the dipole size and $Y$ is the evolution rapidity. In turn, the small-$x$ dynamics of the dipole amplitudes is given by the rcBK equation:
\begin{equation}
  \frac{\partial {\cal N}_{F(A)}(r,Y)}{\partial Y}=\int d^2{\bf r_1}\,
  K^{{\rm run}}({\bf r},{\bf r_1},{\bf r_2}) \left[{\cal N}(r_1,Y)+{\cal N}(r_2,Y)-{\cal N}(r,Y)-
    {\cal N}(r_1,Y)\,{\cal N}(r_2,Y)\right]\,.
\label{bk1}
\end{equation}
For simplicity, we have omitted the subscripts $F(A)$ in the r.h.s of \eq{bk1}.
Using Balitsky's prescription \cite{Balitsky:2006wa}, the kernel in \eq{bk1} reads
\begin{equation}
  K^{{\rm run}}({\bf r},{\bf r_1},{\bf r_2})=\frac{N_c\,\alpha_s(r^2)}{2\pi^2}
  \left[\frac{r^2}{r_1^2\,r_2^2}+
    \frac{1}{r_1^2}\left(\frac{\alpha_s(r_1^2)}{\alpha_s(r_2^2)}-1\right)+
    \frac{1}{r_2^2}\left(\frac{\alpha_s(r_2^2)}{\alpha_s(r_1^2)}-1\right)
  \right]\,,
\label{kbal}
\end{equation}
where ${\bf r_2}={\bf r}-{\bf r_1}$ (throughout the paper we shall use notation $v\equiv |{\bf v}|$ for two-dimensional vectors).

Following \cite{Albacete:2009fh}, we regulate the running coupling in Eqs. (\ref{bk1}) and
(\ref{kbal}) by freezing it to a constant value $\alpha_s^{fr}=0.7$ in the infrared.
A detailed discussion about the different prescriptions proposed to define the running coupling kernel and of the numerical method to solve the rcBK equation can be found in
\cite{Albacete:2007yr}. The only piece of information left to fully complete all the ingredients in \eq{hyb} are the initial conditions for the evolution of the dipole-nucleus(proton) amplitude. Similar to previous works, we take them from the McLerran-Venugopalan (MV) model \cite{McLerran:1997fk}:
\begin{equation}
\mathcal{N}_{F}(r,Y=0)=1-\exp\left[ -\frac{r^2\,Q_{s0}^2}{4}\,\ln\left(\frac{1}{\Lambda\,r}+e\right)\right]\ ,
\end{equation} 
where $Q_{s0}^2$ is the initial saturation scale (probed by quarks), and we take $\Lambda=0.241$ GeV. Contrary to studies of e+p data, we have discarded initial conditions {\it a la} Golec-Biernat-W\"usthoff \cite{Golec-Biernat:1998js}, since their Fourier transform result in an unphysical exponential fall-off of the ugd, and therefore of the hadron spectra as well, at large transverse momenta. Finally, in the large-$N_c$ limit which we have implicitly assumed in order to use the rcBK equation, the gluon dipole scattering amplitude can be expressed in terms of the quark amplitude as
\begin{equation}
\mathcal{N}_A(r, Y ) = 2\, \mathcal{N}_{F} (r, Y)-\mathcal{N}_F^{\,2} (r,Y)\ .
\end{equation}

\begin{figure}[t]
\begin{center}
\includegraphics[height=5.8cm]{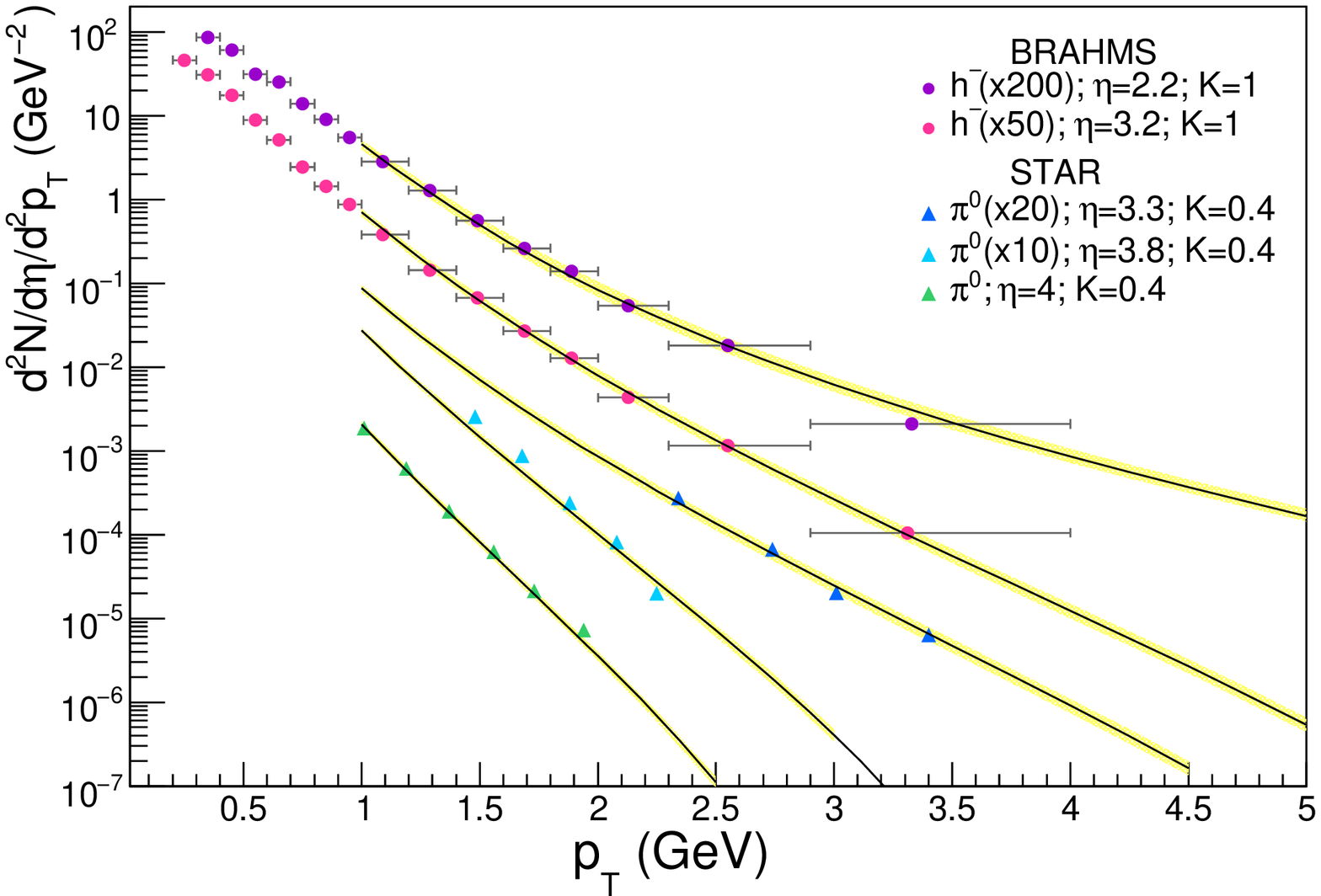}
\includegraphics[height=5.8cm]{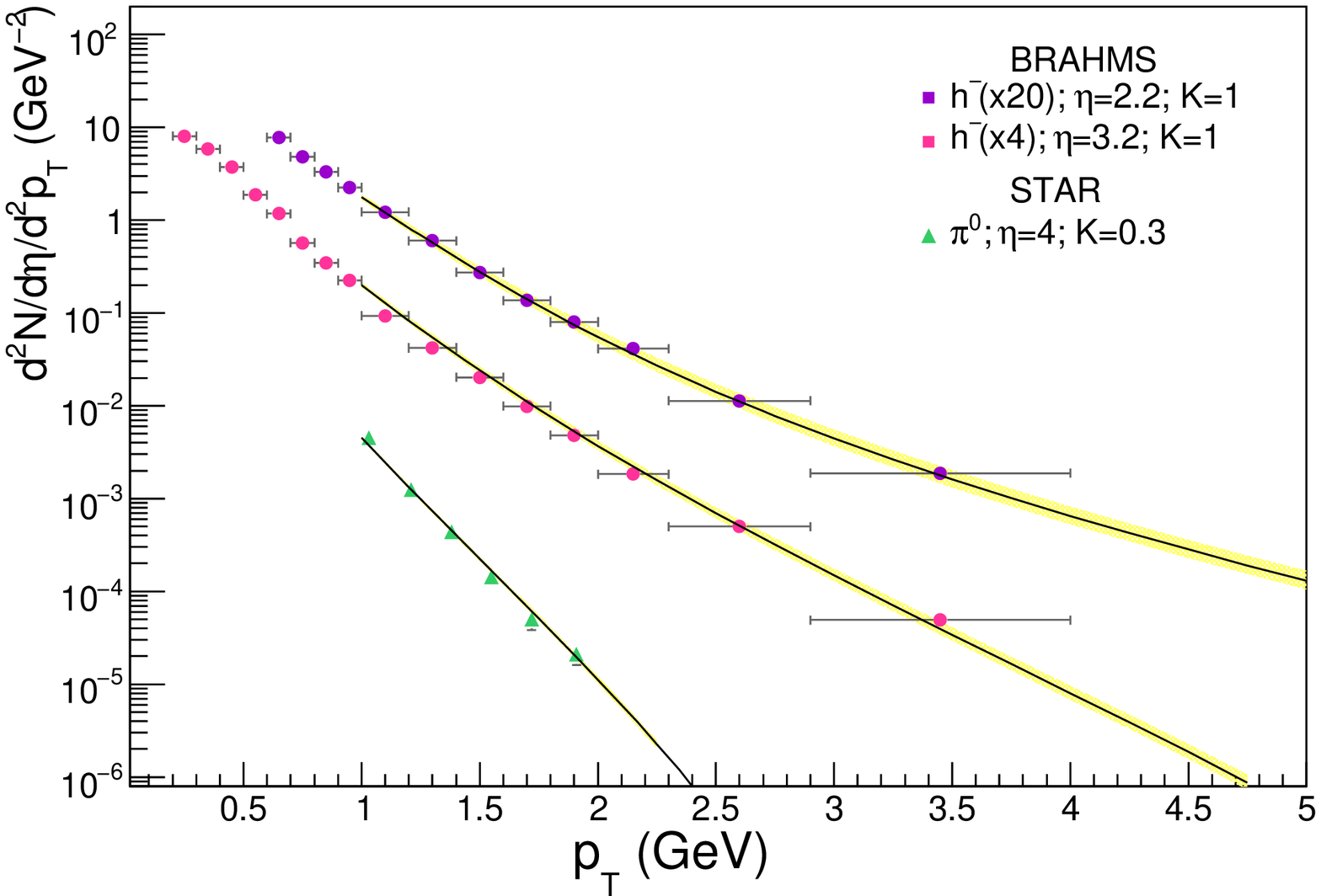}
\end{center}
\caption{Negatively charged hadron and $\pi^0$ yields in proton-proton (at pseudo-rapidities (2.2, 3.2) and (3.3, 3.8 and 4)) and deuteron-gold (at pseudo-rapidities (2.2, 3.2) and 4) collisions at $\sqrt{s_{NN}}=200$ GeV. Data by the BRAHMS and STAR collaborations.}
\label{spp}
\end{figure}

With this setup, we obtain a very good description of RHIC data. \fig{spp} shows the comparison of our results with data for the invariant yield of different hadron species in p+p and d+Au collisions at $\sqrt{s_{NN}}=200$ GeV and rapidities $y_h=2.2$ and 3.2 for negative-charge hadrons (data by the BRAHMS collaboration \cite{Arsene:2004ux}) and $y_h=3.3$, 3.8 and 4 for neutral pions (data by the STAR collaboration \cite{Adams:2006uz}). The only free parameters adjusted to the d+Au data are $x_0,$ the value of $x$ which indicates the start of the small$-x$ evolution, and $Q_{s0}$, the value of the saturation scale at $x=x_0.$ For the gold nucleus we obtain a quark saturation scale $Q^2_{s0}=0.4$ GeV$^2$ at $x_0=0.02$. Values of $x_0$ between 0.015 and 0.025 are allowed within error bands, they are used to generate the yellow uncertainty band in \fig{spp}. A few comments are in order. First, the parameters $Q_{s0}$ and $x_0$ are obtained from minimum-bias data, and therefore $Q^2_{s0}$ should be considered as an impact-parameter averaged value, the saturation scale at the center of the nucleus is bigger. We remind the reader that the corresponding gluon saturation scale is larger, $Q_{s0}^{2,gluon}\!=\!0.9$ GeV$^2$. Second, $Q_{s0}$ and $x_0$ are compatible with other values extracted from e+A \cite{Dusling:2009ni} or A+A \cite{Albacete:2007sm} data. They can be compared with $Q^2_{s0}=0.2$ GeV$^2$ at $x_0=0.007$ obtained in the case of the proton (in \cite{Albacete:2009fh}, $x_0=0.01$ was obtained with $Q_{s0}^2=0.2$ GeV$^2$).
Finally, no K-factor is needed in order to reproduce the charged hadron yields (i.e. $K\!=\!1$), whereas a rapidity independent K-factor $K\!=\!1/3$ is needed to describe the neutral pion data. Although the precise values of the K-factors do not have much meaning due to the 15\% normalization uncertainties of the data, we do not have a good understanding of the strong hadron species dependence.

\section{Nuclear modification factors in p+Pb and Pb+Pb collisions at the LHC}

\begin{figure}[t]
\begin{center}
\includegraphics[height=8cm]{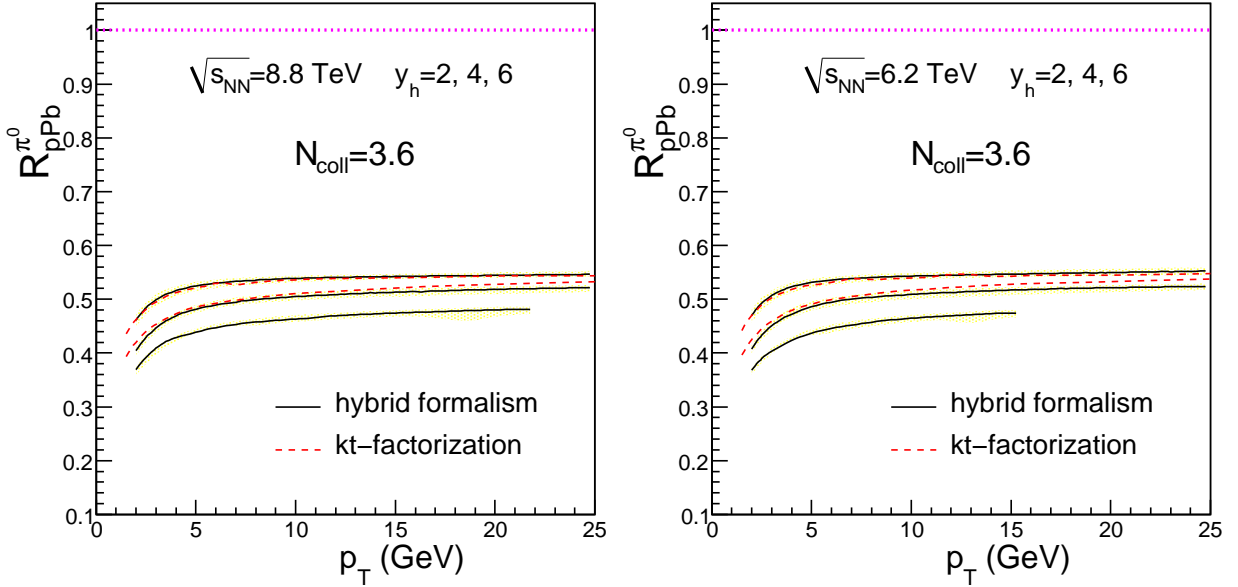}
\end{center}
\caption{Nuclear modification factors for $\pi^0$ production in p+Pb collisions,
$R_{pPb}^{\pi^0}$, for collision energies $\sqrt{s_{NN}}=8.8$ (left) and 6.2 TeV (right) and for rapidities $y_h=2$, 4, and 6. For comparison, the red dashed line corresponds to the same quantity calculated in the k$_t$-factorization scheme.}
\label{rpi0}
\end{figure}

\begin{figure}[t]
\begin{center}
\includegraphics[height=8cm]{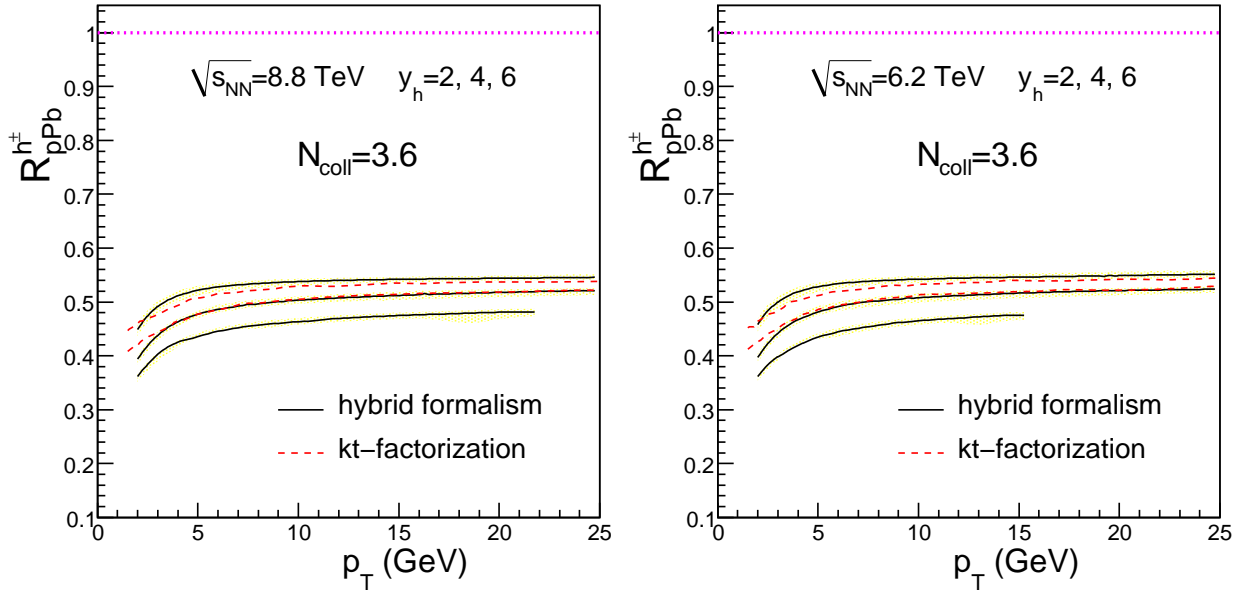}
\end{center}
\caption{Nuclear modification factors for $h^\pm$ production in p+Pb collisions,
$R_{pPb}^{h^\pm}$, for collision energies $\sqrt{s_{NN}}=8.8$ (left) and 6.2 TeV (right) and for rapidities $y_h=2$, 4, and 6. For comparison, the red dashed line corresponds to the same quantity calculated in the k$_t$-factorization scheme.}
\label{rhpm}
\end{figure}

It is straigthforward to use formula \eq{hyb} to calculate forward particle production in p+p and p+Pb at the LHC. We shall present our LHC results in terms of the nuclear modification factor
\begin{equation}
R_{pPb}=\frac{1}{N_{\mbox{coll}}}\frac{dN^{pPb}_h}{dy_h\,d^2p_t}
\Big/\frac{dN^{pp}_h}{dy_h\,d^2p_t}
\label{rpPb}\end{equation}
where $N_{\mbox{coll}}$ is the number of binary proton-nucleon collisions in the p+Pb collision.
In our predictions for p+Pb collisons at the LHC we use $N_{\mbox{coll}}=3.6$, which is half the number of collisions determined in minimum bias d+Au collisions at RHIC \cite{Arsene:2004ux}. Thus, in order to compare our results with experimental data one should renormalize our curves in Figs (\ref{rpi0}) and (\ref{rhpm}) to the number of collisions determined experimentally at the LHC.
Note that computing the ratio \eq{rpPb} removes the sensitivity to the K factors. 
Our $R_{pPb}$ calculations are displayed for two possible LHC energies ($\sqrt{s_{NN}}=8.8$ and 6.2 TeV) in \fig{rpi0} for pion production and \fig{rhpm} for charged hadron production. In both cases, one observes the expected trends that $R_{pPb}$ decreases with increasing $y_h,$ and increases with increasing $p_t.$ Our results for  $R_{pPb}$ indicate that a significant suppression $\sim 1/2$ should be expected already at not too forward rapidities. It is highly likely that the CGC dynamics studied here would also lead to a similar suppression at mid-rapidity (see, e.g. \cite{Tuchin:2007pf}). However, to make a clear quantitative prediction for mid-rapidity one should ensure a proper treatment of high-$x$ effects in the target, which is beyond the scope of this paper.

We also compare the $y=2$ and $4$ curves with predictions obtained with the k$_t-$factorization formalism, in order to check the validity of that approach, and especially to test up to what value of $y$ it can be used. The k$_t-$factorization formula (see \eq{ktfact} below) is valid when the dominant contributions to the cross section come from small values of $x,$ for both the projectile ($x_1\ll 1$) and the target ($x_2\ll 1$). For instance, it only includes gluonic degrees of freedom. This approach is clearly insufficient at very forward rapidities or large $p_t$, where valence quarks of the projectile are important ($x_1\lesssim 1$). However, as can be seen in Figs. \ref{rpi0} and \ref{rhpm}, both formalisms give comparable results, as the lines from k$_t$-factorization overlap with the uncertainty bands spanned by the results from the hybrid formalism. This seems to identify a kinematical window where both approximations (\eq{hyb} and \eq{ktfact} below supplemented with parton fragmentation) are valid. To some extent it is not surprising that both formalisms yield comparable nuclear modification factors, since the suppression is ultimately rooted in the udg's themselves. 

The advantage of the k$_t$-factorization formalism is that it can be used at mid rapidities, when $x_1$ also becomes very small, invalidating formula \eq{hyb}. This is true at the LHC where at mid-rapidity $x_1\sim x_2 \ll 1,$ but not at RHIC where both values of $x$ are generally too large at $y=0.$ Indeed one should keep in mind that $p_t\ e^{y_h}/\sqrt{s_{NN}}$ and $p_t\ e^{-y_h}/\sqrt{s_{NN}}$ are only lower values for $x_1$ and $x_2$ respectively, but that through fragmentation larger values of $x$ actually contribute more.

The k$_t$-factorization formula to describe inclusive gluon production reads \cite{Kovchegov:2001sc}
\begin{equation}
\frac{dN^{AB\to gX}}{d\eta\, d^2 p_t}=\frac{C_F}{\pi}\frac{\alpha_s}{p_t^2}\int d^2bd^2q \,\,
\varphi_A(x_A,q,b)\,\varphi_B(x_B,p_t-q,B_t-b)\,,
\label{ktfact}
\end{equation}
where $B_t$ is the impact parameter of the collision. Gluon fragmentation is not explicitely written down (therefore $x_A=p_t e^y/\sqrt{s_{NN}}$ and $x_B=p_t e^{-y}/\sqrt{s_{NN}}$), but it should also be accounted for.
The unintegrated gluon distribution in \eq{ktfact}, $\varphi$, is actually simply related to the one in \eq{hyb}:
\begin{equation}
\varphi(x,k,b)=\int d^2{\bf r}\,e^{-i{\bf k}\cdot{\bf r}}\,\nabla^2_{\bf r}\,\mathcal{N}(r,Y\!=\!\ln(x_0/x),b)
=k^2\,\tilde{N}(x,k,b)\ .
\end{equation}
A detailed discussion about the definition and physical interpretation of the different udg's discussed here can be found in \cite{Jalilian-Marian:2005jf}.

We had not specified the impact parameter dependence of the ugd's before because it is not needed in a p+p or p+A collision. Indeed in these cases one can write $\int d^2b\ \varphi_p(b)\ \varphi_B(B_t-b)\simeq\varphi_B(B_t)\int d^2b\ \varphi_p(b).$ The $b$-integrated proton udg does not appear in formula \eq{hyb}, it is rather the standard p.d.f.s that describe the (dilute) proton content in this formalism, while in the k$_t$-factorization case the $b$ dependence of $\varphi_p(b)$ can be safely neglected if we are only looking at ratios such as
\eq{rpPb}.
As for the collision impact parameter $B_t$ dependence of the target ugd $\varphi_B$, we have been dealing with minimum-bias data, therefore as mentioned before, the ugd's obtained with
$Q^2_{s0}=0.4$ GeV$^2$ for the gold nucleus (and 0.2 GeV$^2$ for the proton) should be thought of $B_t$ averaged udg's, but in principle we could also look at different centrality bins.
\begin{figure}[t]
\begin{center}
\includegraphics[height=9cm]{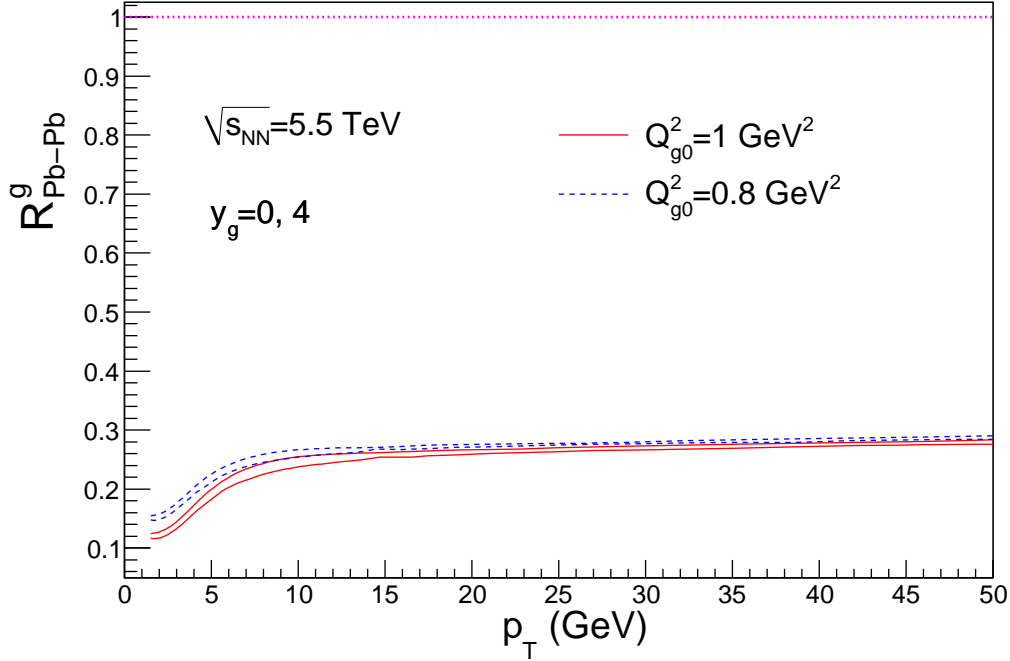}
\end{center}
\caption{Gluon level predictions from k$_t$ factorization for Pb+Pb collisions for rapidities $y=0,4$. Solid lines correspond to an initial gluon saturation scale $Q_{s0}^{gluon\,2}=1$ GeV$^2$, and the dashed ones to $Q_{s0}^{gluon\,2}=0.8$ GeV$^2$.}
\label{rglu}
\end{figure}
Note that formula \eq{ktfact} has only been proven to be valid for p+p and p+A collisons (or {\it dilute-dense} scattering)
\cite{Dumitru:2001ux,Kovner:2001vi,Kovchegov:2001sc,Blaizot:2004wu,Marquet:2004xa}, and there are several hints that it does not hold for A+A collisions (or {\it dense-dense} scattering)
\cite{Gelis:2008rw}, even if there were no final-state effects. However, numerical results seem to indicate that $k_t-$factorization breaking may not be too important in practice
\cite{Albacete:2007sm,Lappi:2008eq,Marquet:2008ue}, but this could be process dependent.
With the above remarks in mind, we shall use \eq{ktfact} to calculate the initial state effects on particle production at mid-rapidity in Pb+Pb collisions at the LHC.

We deal with the impact parameter in the following way: we assume that it factorizes as $\varphi_A(x,k,b)=T_A(b)\tilde\varphi_A(x,k)$ where $T_A(b)$ could be for instance the Woods-Saxon profile and is normalized as $\int d^2b\ T_A(b)=A$. Doing so yields an impact parameter independent $R_{AA},$ as the $b$ integral in \eq{ktfact} is canceled by the number of collisions:
\begin{equation}
R_{AA}=\frac{\int d^2q \,\,\tilde\varphi_A(x_A,q)\,\tilde\varphi_A(x_B,p_t-q)}
{\int d^2q \,\,\tilde\varphi_p(x_A,q)\,\tilde\varphi_p(x_B,p_t-q)}\ .
\label{raa}\end{equation}
This is acceptable for minimum bias results, and in this case the functions $\tilde\varphi_A$ and $\tilde\varphi_p$ are simply related to the averaged $\varphi_A$ and $\varphi_p,$ used for minimun bias p+A and p+p collisions: $N_{\mbox{coll}}^{pA}\ \tilde\varphi_A/\tilde\varphi_p=\varphi_A/\varphi_p.$ We shall again use
$N_{\mbox{coll}}^{pA}=3.6$ in our LHC calculations.

The nuclear modification factor $R_{PbPb}$ we obtain is shown in \fig{rglu} at the gluon level, it corresponds to gluon production immediately after the collision, i.e at proper times $\tau\!=\!0^+$. Obviously, in order to compare our results with data, one should convolute them with {\it final state} effects due to interactions of the produced gluons with the Quark Gluon Plasma, and with hadronization effects as well. Although this is beyond the scope of this work, our results indicate that a sizable part of suppression expected for single hadron yields at the LHC \cite{Abreu:2007kv} may be due to purely initial state effects. Finally, note that the value of the saturation scale probed by gluons at $x_0=0.02$ is $Q^2_{s0}=0.9$ GeV$^2,$ this is used in \fig{rglu} for minimum-bias predictions, rather a band is generated using $Q^2_{s0}=0.8$ and $1$ GeV$^2$. To study different centrality bins in Pb+Pb collisions, first one would have to improve our approximation for the $b$ dependence of the ugd's.

\section{Conclusions}

In this work we have presented a good description of the hadron yields measured at forward rapidities in p+p and d+Au collisions at RHIC using the hybrid formalism proposed in \cite{Dumitru:2005gt} to describe high-energy {\it dilute-dense} scattering. The main new ingredient in our calculation is the use of the BK equation including running-coupling corrections to describe the Bjorken-$x$ dependence of the nuclear (proton) wave functions. With the two free parameters in our work, $x_0$ and $Q_{s0}$ fitted to RHIC data, we extrapolate to LHC energies without further adjustments and predict the suppression of the different forward hadron yields in p+Pb collisions with respect to p+p collisions. Using a different formalism, k$_t$-factorization, we estimate the contribution of initial state effects to the nuclear modification factor in Pb+Pb collisions at the level of gluon production at the LHC. 

While our results offer an additional indication for the presence of CGC effects in RHIC data, the interest now focuses mostly in calibrating the expectations for the Heavy Ion program at the LHC. Our predictions for the LHC rely on the most up-to-date tools within the CGC formalism (Pomeron-loop corrections have been looked at \cite{Iancu:2006uc}, but are only relevant at asymptotically large energies when the running coupling is included \cite{Dumitru:2007ew,Beuf:2007qa}) and will be useful to confirm the tentative conclusions reached at RHIC and to distinguish between alternative physical scenarios, like those proposed in \cite{Kopeliovich:2005ym,Frankfurt:2007rn}  (a complete set of predictions stemming from different formalisms can be found in \cite{Abreu:2007kv}), where the suppression of forward yields at RHIC is due to the non-eikonal propagation of the leading parton, resulting in energy loss in the forward region. Finally, the proper characterization of gluon production in the early stages (before thermalization) of Pb+Pb collisions at the LHC would serve as crucial input for studies of the medium produced in such collisions, such as hydrodynamic simulations or studies of jet quenching. In the latter case, the suppression predicted here due to initial state effects would add to the final state effects due to the presence of a Quark Gluon Plasma.  

\section*{Acknowledgments}        
The work of Javier L. Albacete is supported by a Marie Curie Intra-European Fellowship (FP7-
PEOPLE-IEF-2008), contract number 236376. Cyrille Marquet is supported by the European Commission under the FP6 program, contract No. MOIF-CT-2006-039860.


\end{document}